\documentclass[aps,preprint,superbib]{revtex4}%
\usepackage{amsfonts}
\usepackage{amsmath}
\usepackage{amssymb}
\usepackage{graphicx}%
\providecommand{\U}[1]{\protect\rule{.1in}{.1in}}

\begin{document}
\title{Observation-assisted optimal control of quantum dynamics}
\author{Feng Shuang, Alexander Pechen, Tak-San Ho and Herschel Rabitz}
\affiliation{Department of Chemistry, Princeton University, Princeton, NJ 08544}
\date{\today}

\begin{abstract}
This paper explores the utility of instantaneous and continuous observations
in the optimal control of quantum dynamics. Simulations of the processes are
performed on several multilevel quantum systems with the goal of population
transfer. Optimal control fields are shown to be capable of cooperating or
fighting with observations to achieve a good yield, and the nature of the
observations may be optimized to more effectively control the quantum
dynamics. Quantum observations also can break dynamical symmetries to increase
the controllability of a quantum system. The quantum Zeno and anti-Zeno
effects induced by observations are the key operating principles in these
processes. The results indicate that quantum observations can be effective
tools in the control of quantum dynamics.

\end{abstract}
\maketitle

\section{Introduction}

The control of quantum processes is actively being pursued
theoretically\cite{Rice00,Rabitz0364,Shapiro03} and
experimentally\cite{Walmsley0343,Brixner011} with a variety of control
scenarios\cite{Rabitz884950,Kosloff89201,Shapiro864103,Bergmann905363,
Rice982885}. An increasing number of successful control experiments, including
in complex systems\cite{Gerber98919, Gerber9910381, Kunde00924, Bartels00164,
Brixner0157, Levis01709, Herek02533, Daniel03536}, employ closed-loop optimal
control\cite{Judson921500}. The latter experiments commonly aim to enhance the
yield of a particular desired final state, where a measurement of the quantum
system is only performed after the evolution is over. Utilizing quantum
observations \textit{during} the control process may offer an opportunity to
enhance performance\cite{Rice049984, Sugawara05204115}. Recent
studies\cite{Shuang049270,Shuang06154105} also have shown that controlled
quantum dynamics can operate in the presence of significant field noise and
decoherence, and even cooperate with them under suitable circumstances. This
paper will demonstrate that analogous control cooperation can occur between
the actions of applied external fields and observations with both aiming to
manipulate the system's quantum dynamics.

A characteristic feature of quantum mechanics is that the performance of a
measurement unavoidably affects the subsequent system dynamics. A well known
manifestation of this observation driven back action is the uncertainty
principle\cite{Mensky1993}. A direct influence of a measurement is revealed
through a change in the system state. In the von Neuman view of quantum
mechanics, an instantaneous measurement projects the state of the system onto
an eigenstate of the observable operator\cite{VonNeumann1955}. The measurement
process induces irreversible dynamics and results in a lack of system
coherence, corresponding to the off-diagonal matrix elements of the density
matrix decaying to zero or the phase of the wavefunction amplitudes being randomized.

This paper is concerned with measurements carried out over a period of time.
One of the earliest approaches to continuous quantum measurements was
suggested by Feynman in terms of path integrals\cite{Feynman48367}. When
measurements are performed the Feynman propagator is modified by restricting
the paths to cross (or not to cross) certain space-time regions. An
approximate technique was developed by Mensky\cite{Mensky94159} who
incorporated Gaussian cut-offs in the phase space path integrals and showed
its equivalence to the phenomenological master equation approach for open
quantum system dynamics using models of system-environment
coupling\cite{Walls1994}.

Prevention of a quantum system's time evolution by means of repetitive,
frequent observations or continuous observations of the system's state is
called the quantum Zeno effect (QZE). The QZE was proposed by Misra and
Sudarshan\cite{Misra77756} and was experimentally
demonstrated\cite{Itano902295} in a repeatedly measured two-level system
undergoing Rabi oscillations. A time-dependent observable projection operator
inducing up to $100\%$ transfer from one state to another
state\cite{Roy004019} is called the quantum anti-Zeno effect (QAZE). The
impacts of QZE and QAZE operations are the key processes explored in this
paper to help control quantum dynamics.

This paper explores the scope of what might be gained in terms of better
control performance from utilizing suitable observations. The practical means
of executing observations in this fashion will be the subject of future works.
The remainder of the paper is broken down the following way. Section II
reviews the main concepts of performing instantaneous and continuous
measurements, which are utilized in this paper. Section III presents the model
system, and Section IV presents simulations of the closed-loop management of
quantum dynamics assisted by measurements. A brief summary of the findings is
given in Section V.

\section{Quantum Observations}

\subsection{Instantaneous observations}

An ideal instantaneous measurement occurs at one point of time, or a sequence
of such observations can follow each other at different
times\cite{VonNeumann1955}. An instantaneous measurement may be characterized
by a set of projectors $\left\{  P_{i}\right\}  $ satisfying conditions of
completeness and orthogonality%
\begin{equation}
\sum_{k}P_{k}=1,\ \ P_{i}P_{j}=0\ \text{for }i\neq j\text{.}%
\end{equation}
The instantaneous measurement converts the state $\rho$ into the state
$\rho^{\prime}$,%
\begin{equation}
\rho^{\prime}=\sum_{k}P_{k}\rho P_{k}\text{.} \label{MRho}%
\end{equation}
We may also observe a physical quantity represented by the operator $A$,
\begin{equation}
A=\sum_{i}a_{i}\left\vert a_{i}\right\rangle \left\langle a_{i}\right\vert
\text{,}%
\end{equation}
where $a_{i}$ and $\left\vert a_{i}\right\rangle $ are the i-th eigenvalue and
eigenstate, respectively, of the observable operator $A$, and the density
matrix maybe expressed in the form%
\begin{equation}
\rho=\sum_{k,j}\rho_{kj}\left\vert a_{k}\right\rangle \left\langle
a_{j}\right\vert \text{.}%
\end{equation}
When a measurement of $A$ is performed, the reduction
\begin{equation}
\rho_{kj}\rightarrow0,\text{ for }a_{k}\neq a_{j}%
\end{equation}
occurs, thereby destroying the coherence between nondegenerate states of
operator $A$. If $A$ has no degenerate eigenstates, then $\rho$ will contain
only diagonal elements after an instantaneous quantum measurement%
\begin{equation}
\rho\rightarrow\sum_{k}\rho_{kk}\left\vert a_{k}\right\rangle \left\langle
a_{k}\right\vert \text{.} \label{InsM}%
\end{equation}
If a projection operator $P$ is observed, it's easy to deduce from Eq.
(\ref{MRho}) that after the observation process, the density matrix is
transformed to $\rho^{\prime}$ given by%

\begin{subequations}
\begin{align}
\rho^{\prime}  &  =P\rho P+\left(  1-P\right)  \rho\left(  1-P\right) \\
&  =\rho-\left[  P,\left[  P,\rho\right]  \right]  \text{.} \label{OInst}%
\end{align}
The operation $\left[  P,\left[  P,\rho\right]  \right]  $ may be viewed as
the "kick" resulting from the instantaneous observation of the projection
operator $P$.

\subsection{Continuous observations}

The employment of restricted path integrals and master equations (ME) form two
equivalent techniques in the theory of continuous quantum
measurements\cite{Mensky94159}. For simplicity, we adopt the ME formalism.
With a continuous measurement of a single observable $A$ the ME takes the form
\cite{Walls1994}:%

\end{subequations}
\begin{equation}
\dot{\rho}=-i\left[  H,\rho\right]  -\frac{1}{2}\kappa\left[  A,\left[
A,\rho\right]  \right]  \text{.} \label{ME}%
\end{equation}
Here, $H$ is the Hamiltonian of the measured system, and $\kappa$ indicates
the "strength" of the observation. Equation (\ref{ME}) is similar to the
equation describing a system interacting with the environment. The first term
in Eq. (\ref{ME}) describes the propagation of the free system, while the
second term provides the decay of the off-diagonal matrix elements, such that%
\begin{equation}
\frac{\partial}{\partial t}\left\langle a_{i}|\rho|a_{j}\right\rangle
=-i\left\langle a_{i}\right\vert \left[  H,\rho\right]  \left\vert
a_{j}\right\rangle -\frac{1}{2}\kappa\left(  a_{i}-a_{j}\right)
^{2}\left\langle a_{i}|\rho|a_{j}\right\rangle \text{.}%
\end{equation}

\section{The Model System}

The effect of measurements on controlled quantum dynamics is explored here in
the context of population transfer in several multilevel systems characterized
by the Hamiltonian,%
\begin{subequations}
\begin{align}
H  &  =H_{0}-\mu E(t)\text{,}\label{Ht}\\
H_{0}  &  =\sum_{v}\varepsilon_{\upsilon}\left\vert \upsilon\right\rangle
\left\langle \upsilon\right\vert \text{,} \label{H0}%
\end{align}
where $\left\vert \upsilon\right\rangle $ is an eigenstate of $H_{0}$ and
$\varepsilon_{\upsilon}$ is the associated field-free eigenenergy, and $\mu$
is the dipole operator. The control field $E(t)$ is taken to have the
following form which may be implemented in the laboratory\cite{Rabitz05013419,
Dantus_OptExpress061030},
\end{subequations}
\begin{subequations}
\label{E0}%
\begin{align}
E(t)  &  =s(t)\sum_{l}^{M}A_{l}\cos\left(  \omega_{l}t+\theta_{l}\right)
\text{,}\label{GE}\\
s(t)  &  =\exp\left[  -\left(  t-T_{f}/2\right)  ^{2}/2\sigma^{2}\right]
\text{,}%
\end{align}
where $\left\{  \omega_{l}\right\}  $ are the $M$ allowed resonant transition
frequencies of the system and $s(t)$ is the pulse envelope function. The
controls are the amplitudes $\left\{  A_{l}\right\}  $ and phases $\left\{
\theta_{l}\right\}  $.

Closed-loop control simulations will be performed to model a laboratory
circumstance with the cost function:%

\end{subequations}
\begin{subequations}
\label{Obj}%
\begin{align}
J\left[  E(t)\right]   &  =\left\vert O\left[  E(t)\right]  -O_{T}\right\vert
^{2}+\alpha F\text{,}\label{J0}\\
F  &  =\sum_{l}\left(  A_{l}\right)  ^{2}\text{,} \label{F0}%
\end{align}
where $O_{T}$ is the target value (expressed as a percent yield) and
\end{subequations}
\begin{equation}
O\left[  E(t)\right]  =\text{Tr}[\rho(T_{f})\hat{O}] \label{O}%
\end{equation}
is the outcome produced by the field $E(t)$ at time $T_{f}$, and $F$ is the
fluence of the control field whose contribution is weighted by the constant,
$\alpha>0$. In the present work, $\hat{O}=\left\vert \Psi_{f}\right\rangle
\left\langle \Psi_{f}\right\vert $ is a projection operator for transferring
population into the target state $\left\vert \Psi_{f}\right\rangle $. The goal
of this study is to explore the role that observations can play in aiding the
control process and possibly reducing the fluence of $E\left(  t\right)  $ to
more effectively achieve the desired final state.

\section{Observations Serving as Controls}

In this section, we numerically investigate four simple model systems in Fig.
\ref{Fig_Model} to explore the use of observations in the control of quantum
dynamics. In model 1, the control field is optimized and shown to be capable
of fighting against the effect of instantaneous observations of different
operators when they act as disturbances. The optimal control fields are also
capable of cooperating with the observation of the dipole to attain a better
value for the objective, even when the desired target yield is large. In model
2, the control field is fixed but the instantaneous observed operators are
optimized. It is shown how the presence of even a non-optimal control field
can help the observation processes meet the target yield. Quantum observations
are used to break the dynamical symmetry in model 3, and the optimized
continuous observations are shown to assist in making the control process more
effective. In model 4, continuous observation is used to avoid population loss
into an undesired state. In the first two models, the QAZE is used to induce
population transfer, while the QZE is the operating process in models 3 and 4
used to prohibit population transfer. In all the illustrations a genetic
algorithm\cite{Goldberg97} is employed to optimize the control fields and observations.

\subsection{Model 1}

This model uses the five-level system in Fig. \ref{Fig_Model}(a) with
eigenstates $\left\vert i\right\rangle $, $i=0,\cdots,4$ of the field free
Hamiltonian $H_{0}$, having only nearest neighbor transitions with frequencies
$\omega_{01}=1.511$, $\omega_{12}=1.181$, $\omega_{23}=0.761$, and
$\omega_{34}=0.553$ in rad fs$^{-1}$, and associated transition dipole moments
$\mu_{01}=0.5855$, $\mu_{12}=0.7079$, $\mu_{23}=0.8352$ and $\mu_{34}=0.9281$
in $1.0\times10^{-30}$ C m. The target time is $T_{f}=200$ fs, the pulse width
in Eq. (\ref{E0}) is $\sigma=30$ fs, and the weight coefficient in Eq.
(\ref{J0}) is $\alpha=0.05$. The control objective is to transfer population
from the initially prepared ground state $\left\vert 0\right\rangle $ to the
highest excited state $\left\vert 4\right\rangle $, such that $\hat
{O}=\left\vert 4\right\rangle \left\langle 4\right\vert $ in Eq. (\ref{O}). As
a reference control case, we first determine the optimal control field without
any observations. Figure \ref{Fig_Field} depicts the amplitude and power
spectrum of control field. A population transfer of $98.44\%$ is achieved in
the target state by the optimal control field which has the fluence $0.063$.
The fields in all of the illustrations in this paper have general structure
similar to that in Fig. \ref{Fig_Field} due to the imposed form in Eq.
(\ref{E0}), and these other fields will not be explicitly shown.

Assuming that for some auxiliary purpose we need to detect a physical quantity
$A$ at the middle of dynamical evolution at%
\begin{equation}
T_{m}=\frac{T_{f}}{2}\text{,} \label{Tm}%
\end{equation}
Table I shows how the optimally determined control fields (i.e., each
observation has a distinct optimal field of the form in Eq. (\ref{E0})) fight
against the observation of the dipole $\mu$, the energy $H_{0}$ and the
population of each level%
\begin{equation}
P_{k}=\left\vert k\right\rangle \left\langle k\right\vert \label{PopM}%
\end{equation}
with $k=0,\cdots,4$. The second column of Table I indicates that the control
field can fight very effectively with the disturbance caused by the individual
quantum observations. Note that the results for population observations (the
third column of Table I with $P_{k}$, $k=0,\cdots,4$) are all near zero, which
reveals the mechanism employed by each control field to fight against its
associated observation: the control field $E_{k}\left(  t\right)  $ associated
with the observation operator $P_{k}$ drives the system to a state
$\rho\left(  T_{m}\right)  =\left\vert \psi\right\rangle \left\langle
\psi\right\vert $ that is nearly orthogonal to the observed state $\left\vert
k\right\rangle $,
\begin{equation}
\left\langle \psi|k\right\rangle \approx0\text{,}%
\end{equation}
such that
\begin{equation}
\left[  P_{k},\left[  P_{k},\rho\left(  T_{m}\right)  \right]  \right]
\approx0\text{.}%
\end{equation}
This behavior assures that the observation of $P_{k}$ has little effect on the
system state, or equivalently the "kick" from the observation disappears from
Eq. (\ref{OInst}). After checking the results of observing the energy and
dipole, we find a similar mechanism: their observed values at $T_{m}$ are all
nearly equal to an eigenvalue of the observed operators, which means that the
control field drives the system to an eigenstate of the observed operators at
$T_{m}$, again so that the observation has little effect on the system state.
It is evident in this case that the deleterious impact of any instantaneous
observation can be corrected because a suitable control field can drive model
1 to any state. The fourth column in Table I uses the optimal fields
determined in the presence of the observation, but the dynamics are carried
out in the end without the observation being present. The very similar yields
in the second and fourth columns are consistent with the mechanism indicated
above. The last column in Table I shows that fighting against the disturbance
created by the observation increases the control field fluence, whose values
depend on the particular observation operator. These results collectively
indicate that in the present model when seeking a high target yield the most
efficient strategy for the control field is to fight the impact of the
observation, which is acting as a disturbance disruptive to the control goal.

The observation of the dipole can have the dual competitive role of destroying
the coherence of the system, while also inducing population transfer. A
calculation shows that performance of an observation of the dipole $\mu$
without the control field being present can induce $22.19\%$ population
transfer from the initial state to the target state. Table II describes how
the optimal control fields work with an observation of the dipole $\mu$ to
reach different posed target yields. The second column shows that the target
yield can be reached in all the cases, with some lose in achieved fidelity at
the highest demanded yield of $O_{T}=100\%$. In order to reveal the
contributions of the observations upon the optimally controlled dynamics, the
third column of Table II shows the yield from the field alone without the
observation being made, yet with the field determined in the presence of the
observation. Comparison of the second and third column in Table II shows that
a remarkable degree of cooperation is found when the expected target yield
lies in the range greater than $22.19\%$ up to $\sim50\%$, and the effect is
even evident at the $70\%$ target yield. For example, at the target yield of
$O_{T}=40\%$, the observation and optimal field alone, respectively, produce
yields of $22.19\%$ and $2.69\%$. But, the same field operating in the
presence of the observation produces a yield of $39.82\%$. This behavior
indicates that the field is cooperating with the observation to more
effectively achieve the posed goal. Above a target yield of $\sim80\%,$ the
field works to fight against the observation acting as a disturbance. The
fourth column of Table II shows that the fluence generally follows this
behavior. Below a target yield of $\sim70\%$ and higher than $22.19\%$, the
reduced fluence with the observation being present shows the enhanced control
efficiency. Above that value the observation increasingly acts as a
disturbance, which calls for an enhanced field fluence to fight against it.

\subsection{Model 2}

Model 2 has the same Hamiltonian and dipole elements as model 1, but we
concentrate on studying the effects of a sequence of instantaneous
observations treated now as controls for the population transfer. Again, the
objective is to transfer population from level $0$ to level $4$ at the target
time $T_{f}=200$ fs. We assume that any projection operator may be observed in
a suitably performed experiment. A sequence of $N$ instantaneous projection
observations, specified by the operators
\begin{subequations}
\label{ProOp}%
\begin{align}
P_{k} &  =\left\vert \psi_{k}\right\rangle \left\langle \psi_{k}\right\vert
\text{, }k=1,\cdots,N\text{,}\\
\left\vert \psi_{k}\right\rangle  &  =\sum_{j=0}^{4}a_{jk}\left\vert
j\right\rangle \text{, \ }\sum_{j=0}^{4}\left\vert a_{jk}\right\vert ^{2}=1
\end{align}
are performed at equally spaced time intervals,
\end{subequations}
\begin{equation}
t_{k}=\frac{k}{N+1}T_{f}\text{, }k=1,\cdots,N\text{,}%
\end{equation}
respectively. The variables subjected to optimization are the complex
coefficients $\left\{  a_{jk}\right\}  $ in the projection operators of Eq.
(\ref{ProOp}). A control field, of the form Eq. (\ref{E0}), is utilized in
some of the simulations, but the amplitudes and phases are picked \textit{a
priori} without any attempt at optimization. At first, the control field is
turned off and the objective functional,%
\begin{equation}
J\left[  \mathbf{P}_{N}\right]  =\left\vert O\left[  \mathbf{P}_{N}\right]
-100\%\right\vert ^{2}\text{,}\label{JOb}%
\end{equation}
is optimized with respect to the coefficients $\left\{  a_{jk}\right\}  $ in
the $N$ observed operators
\begin{equation}
\mathbf{P}_{N}=\left(  P_{1},\cdots,P_{N}\right)  \text{.}%
\end{equation}
In Eq.(\ref{JOb}) $O\left[  \mathbf{P}_{N}\right]  $ is the population yield
attained from the observations without the control field. The second column of
Table III shows the largest attainable population transfer with different
numbers of optimized observations when the control field is off. It has been
proved that the QAZE induced by suitable time-dependent measurements can fully
transfer population to a target state in the frequent measurement
limit\cite{Roy004019}, $\lim_{N\rightarrow\infty}O\left[  \mathbf{P}%
_{N}\right]  =100\%$. We now introduce a weak control field of the form in Eq.
(\ref{E0}) with all of the amplitudes being $0.07$ and phases set at $0.0$.
The target time is $T_{f}=200$ fs, and the pulse width in Eq. (\ref{E0}) is
$\sigma=30$ fs. This fixed non-optimal control field can only drive $12.93\%$
of the population to target state when acting alone (i.e., without
observation). The objective is now a functional of both the control field and
measured operators,
\begin{equation}
J\left[  E\left(  t\right)  ,\mathbf{P}_{N}\right]  =\left\vert O\left[
E(t),\mathbf{P}_{N}\right]  -100\%\right\vert ^{2}\text{,}%
\end{equation}
but still only the observation operators $\mathbf{P}_{N}$ are optimized. The
third column in Table III shows the attained population transfer induced by
both the control field and the optimized observations acting together. \ The
contribution from the observations acting alone is listed in the fourth
column. A high degree of cooperation between the control field and observation
is found. For example, for $N=5$ the observations carried out alone produce a
yield of $20.46\%$ and the yield from the non-optimal control field alone is
$12.93\%$, but the yield from both acting together is $79.22\%$, much larger
than their simple summation. Table III indicates that when $N<9$, the presence
of the control field is helpful for achieving a higher yield. Further
numerical simulations show that, when $N\geq9$, the presence of the control
field becomes less helpful, which reflects the strength of observations acting
alone as controls. This behavior may be confirmed by an analytical
assessment\cite{Pechen06052102, Shuang2006_1} of $O\left[  P_{N}\right]  $,
which proves that, when $N\geq9$, the maximum population transfer induced by
$N$ observations is larger than $80\%$.

\subsection{Model 3}

Model 3 in Fig. \ref{Fig_Model}(b) is a high symmetry three-level system with
the Hamiltonian $H_{0}$ and dipole $\mu$ given by
\begin{equation}
H_{0}=\left(
\begin{array}
[c]{ccc}%
1 & 0 & 0\\
0 & 2 & 0\\
0 & 0 & 3
\end{array}
\right)  \text{, }\mu=\left(
\begin{array}
[c]{ccc}%
0 & 1 & 0\\
1 & 0 & 1\\
0 & 1 & 0
\end{array}
\right)  \text{. }%
\end{equation}
The system is initially prepared in its ground state $\left\vert
0\right\rangle $, and the objective is to transfer the population to state
$\left\vert 1\right\rangle $ at target time $T_{f}=200$ fs. If only a
dipole-coupled external field is employed, the high symmetry in $H_{0}$ and
$\mu$ implies that the system is not fully controllable, and by inspection at
most $50\%$ of the population maybe be transferred to state $\left\vert
1\right\rangle $. This assessment can be made rigorous in the following
analysis. It has been proved\cite{Rabitz011} that there is a hidden dynamical
symmetry in this system,
\begin{equation}
\left\vert C_{0}\left(  t\right)  C_{2}\left(  t\right)  -\frac{C_{1}%
^{2}\left(  t\right)  }{2}\right\vert =\left\vert C_{0}\left(  0\right)
C_{2}\left(  0\right)  -\frac{C_{1}^{2}\left(  0\right)  }{2}\right\vert
=0\text{,} \label{Ck}%
\end{equation}
where $C_{k}\left(  t\right)  \,,$ $k=1,2,3$ are complex coefficients of the
system wavefunction%
\begin{equation}
\psi\left(  t\right)  =\sum_{k=0}^{2}C_{k}\left(  t\right)  \left\vert
k\right\rangle \text{.}%
\end{equation}
Rewriting Eq. (\ref{Ck}) in terms of density matrix elements gives%
\begin{equation}
\rho_{00}\left(  t\right)  \rho_{22}\left(  t\right)  =\frac{\rho_{11}%
^{2}\left(  t\right)  }{4}\text{.} \label{RhoS}%
\end{equation}
The following inequality based on Eq. (\ref{RhoS}) shows that no more than
$50\%$ of the population can be driven from its ground state $\left\vert
0\right\rangle $ to the state $\left\vert 1\right\rangle $%
\begin{equation}
\rho_{11}\left(  t\right)  =2\sqrt{\rho_{00}\left(  t\right)  \rho_{22}\left(
t\right)  }\leq\rho_{00}\left(  t\right)  +\rho_{22}\left(  t\right)
=1-\rho_{11}\left(  t\right)  \text{.}%
\end{equation}
To explore if observations can break the $50\%$ yield limit, first a simple
instantaneous observation and then a time-dependent continuous observation is
applied. The control field is a simple resonant rectangular
pulse\cite{Richardson_JLT010501},%
\begin{equation}
\left\{
\begin{array}
[c]{c}%
E\left(  t\right)  =A\cos t\text{, }0\leq t\leq T_{f}\text{,}\\
E\left(  t\right)  =0\text{, \ \ \ \ \ \ \ otherwise, \ }%
\end{array}
\right.  \label{EA}%
\end{equation}
where only the amplitude $A$ is adjusted for optimization.

First, an instantaneous observation is performed at the middle of the control
$T_{m}=T_{f}/2$. Table IV shows various control yields when different
instantaneous observations are carried out, where $P_{k}$ is the population
measurement operator in Eq. (\ref{PopM}). The simulation shows that an
instantaneous population observation of state $\left\vert 0\right\rangle $ or
$\left\vert 2\right\rangle $ can increase the population transfer to the
target state $\left\vert 1\right\rangle $, but at the expense of requiring
stronger control fields. In contrast, an observation of the target state
population is not helpful. This behavior can be explained by the broken
dynamical symmetry induced by the observation of state $\left\vert
0\right\rangle $ or $\left\vert 2\right\rangle $, but this outcome will not be
the case from observation of state $\left\vert 1\right\rangle $. An analytical
treatment\cite{Shuang2006_1} shows that the maximum attainable population
transfer to the level $\left\vert 1\right\rangle $ by a coherent field
assisted from measuring $P_{0}$ or $P_{2}$ is $68.7\%$, which is closely
approximated by the value of $\simeq67\%$ in Table IV.

Now consider carrying out time-dependent continuous observations together with
a control field $E\left(  t\right)  $ having the form in Eq. (\ref{EA}), where
the density matrix satisfies%
\begin{equation}
\dot{\rho}=-i\left[  H_{0}-\mu E\left(  t\right)  ,\rho\right]  -\frac{1}%
{2}\kappa\left(  t\right)  \left[  P,\left[  P,\rho\right]  \right]  \text{.}%
\end{equation}
Here the observation strength $\kappa\left(  t\right)  $ is allowed to be
time-dependent, and a simple form of $\kappa\left(  t\right)  $ is adopted as
it proved to be sufficient in the control of model 3:
\begin{equation}
\kappa\left(  t\right)  =\left\{
\begin{array}
[c]{l}%
\gamma\text{,}\ \text{ }T_{1}<t<T_{2}\text{,}\\
0\text{,}\ \text{ otherwise. \ \ \ \ \ \ \ }%
\end{array}
\right.
\end{equation}
In this case the objective functional $J$ is optimized with respect to not
only the control field parameter $A$ in Eq. (\ref{EA}), but also the
observation strength $\gamma$ and time interval $T_{1}$, $T_{2}$,
\begin{equation}
J\left[  A,\gamma,T_{1},T_{2}\right]  =\left\vert O\left[  A,\gamma
,T_{1},T_{2}\right]  -100\%\right\vert ^{2}+\alpha A^{2}\text{.} \label{JMF}%
\end{equation}
The coefficient $\alpha$ in Eq. (\ref{JMF}) is $0.01$. In the simulation, the
observation strength $\gamma$ was optimized over the range from $0.0$ to
$5.0$. Table V shows that with the help of the optimized continuous
observations of the population in state $\left\vert 0\right\rangle $ or
$\left\vert 2\right\rangle $, the control field can induce almost $100\%$
population transfer between the initial state $\left\vert 0\right\rangle $ and
target state $\left\vert 1\right\rangle $. As expected, observation of the
state $\left\vert 1\right\rangle $ is not helpful. Figure \ref{Fig_Symm}(a)
shows the state populations when the optimized continuous observation is on
state $\left\vert 0\right\rangle $ and Fig. \ref{Fig_Symm}(b) shows the state
populations when the optimized continuous observation is on state $\left\vert
2\right\rangle $. The results in Fig. \ref{Fig_Symm} indicate that the
observation of $P_{0}$ or $P_{2}$ eliminate population from the state being
observed, and the three-level system becomes an effective two-level system in
the time interval $T_{1}<t<T_{2}$. This behavior is consistent with the
observation acting under the QZE. In both cases $\gamma$ adopts its maximum
value of $5.0$ under optimization to evidently take full advantage of the QZE.
The simulations with this simple model show that observations can
fundamentally alter the effective dynamical structure of a quantum system.
This role of observations will be confirmed again in a forthcoming analytical
treatment\cite{Shuang2006_1}. Naturally, for more complex systems, additional
specially tailored time-dependent observations may be required for optimal
impact on the controlled dynamics.

\subsection{Model 4}

\begin{figure}
[h]
\begin{center}
\includegraphics[
trim=0.000000in 0.000000in 0.000000in -0.084618in,
height=4.5164in,
width=6.0222in
]%
{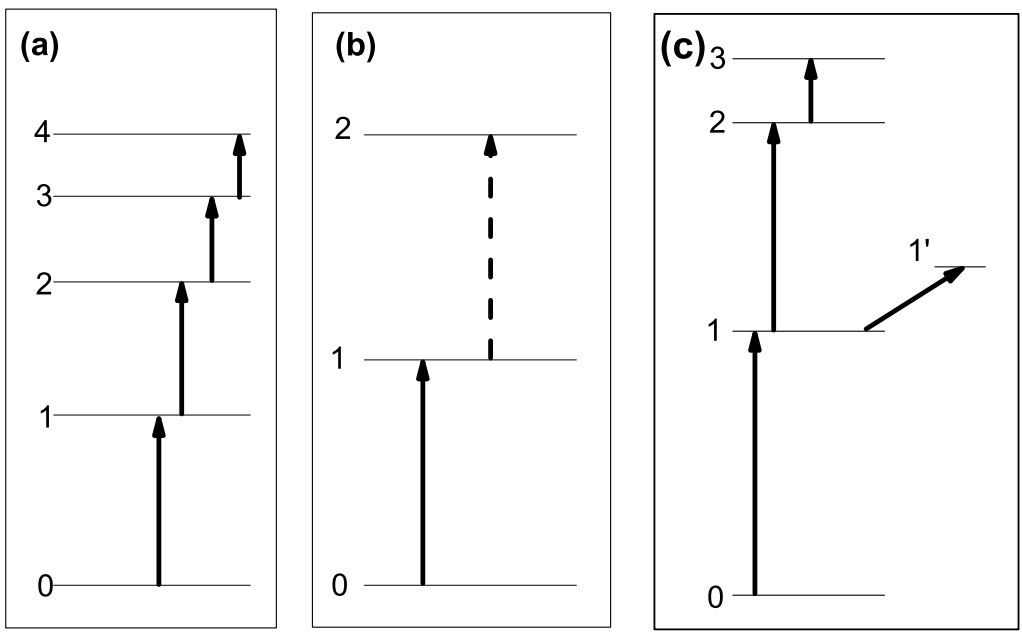}%
\caption{F Shuang et al}%
\label{Fig_Model}%
\end{center}
\end{figure}

The structure of model 4 is given in Fig. \ref{Fig_Model}(c), and the
objective is to transfer population from level $0$ to level $3$. There are two
degenerate transitions, $\omega_{11^{\prime}}=\omega_{23}=0.8$, and the other
transition frequencies are $\omega_{01}=3.3$, $\omega_{12}=2.6$. The non-zero
dipole elements are: $\mu_{01}=0.13$, $\mu_{12}=0.15$, $\mu_{23}=0.23$ and
$\mu_{11^{\prime}}=0.21$. The control field has the form of Eq. (\ref{E0})
with the resonant three amplitudes and phases subjected to optimization. The
target time is $T_{f}=200$ fs, the pulse width in Eq. (\ref{E0}) is
$\sigma=30$ fs and the weight coefficient in Eq. (\ref{J0}) is $\alpha=0.01$.
The simulation in the first row of Table VI shows that under these conditions,
with no observation, the control field can only drive $71.96\%$ of population
to the target state, mainly because some population is locked in the undesired
state $\left\vert 1^{\prime}\right\rangle $. If a constant continuous
observation of the population of state $\left\vert 1^{\prime}\right\rangle $
is carried out, the dynamics of model 4 is described by following equation:%
\begin{subequations}
\begin{align}
\dot{\rho}  &  =-i\left[  H_{0}-\mu E\left(  t\right)  ,\rho\right]  -\frac
{1}{2}\kappa\left[  P_{1^{\prime}},\left[  P_{1^{\prime}},\rho\right]
\right]  \text{,}\label{M4Dyna}\\
P_{1^{\prime}}  &  =\left\vert 1^{\prime}\right\rangle \left\langle 1^{\prime
}\right\vert \text{.}%
\end{align}
Table VI shows that increasing $\kappa$ results in a reduction of the
population in state $\left\vert 1^{\prime}\right\rangle $. The phenomena can
be explained by the QZE: the strong continuous measurement of state
$\left\vert 1^{\prime}\right\rangle $ prohibits population transfer between
state $\left\vert 1\right\rangle $ and $\left\vert 1^{\prime}\right\rangle $
and avoids population loss to the undesired state $\left\vert 1^{\prime
}\right\rangle $, thereby increasing the population in the target state. In
all the cases in Table VI the fluence of the control field remains
approximately the same at F $\simeq$ 0.57, despite the fact that some of the
amplitudes $A_{l}$ in Eq. (\ref{GE}) changed to some degree as $\kappa$
varied. Population loss to undesired states is commonly encountered in the
practical control of quantum dynamics. This model shows a mechanism to avoid
the loss.

\section{\bigskip Conclusions}

\bigskip This paper discusses observations serving as indirect controls in the
manipulation of quantum dynamics. In this context, the field entering the
Hamiltonian can be viewed as a direct control. Instantaneous and continuous
observations were both considered along with control fields to manipulate
population transfer. The simulations show that suitable observations can be
very helpful in the manipulation of quantum dynamics. In favorable cases the
optimal control field can cooperate with observations to achieve the target
more effectively, even when the objective yield is large. In turn, optimal
observations can work with an existing or constrained control field to
transfer more population from an initial state to a target state. Observations
can break dynamical symmetry to increase controllability as well as prohibit
transfer of amplitude to undesired states. The QZE and QAZE are the key
operational processes associated with the observations to assist the control
field to more effectively achieve the target objective. The performance of
optimal observations hopefully will become routine with advancing technology,
as observations can be powerful tools in the control of quantum dynamics.
\end{subequations}
\begin{acknowledgments}
The authors acknowledge support from the NSF, DARPA and ARO-MURI.
\end{acknowledgments}

\bibliographystyle{apsrev}

\pagebreak

\bigskip TABLE I. Optimal control fields fighting against the disturbance of
various observations for model 1 with the goal of a high target yield
$O_{T}=100\%$.%

\begin{tabular}
[c]{|c|c|c|c|c|}\hline
$A\ ^{a}$ & $O\left[  E\left(  t\right)  ,A\right]  \ ^{b}$(\%) & Tr$\left[
\rho\left(  T_{m}\right)  A\right]  \ ^{c}$ & $O\left[  E\left(  t\right)
,0\right]  \ ^{d}$(\%) & $F\ ^{e}$\\\hline
-- & 98.44 & -- & 98.44 & 0.063\\\hline
$\mu$ & 92.42 & 0.66 & 94.03 & 0.37\\\hline
$H_{0}$ & 85.45 & 3.94 & 85.17 & 1.29\\\hline
$P_{0}$ & 97.14 & \textit{0.0037 } & 95.77 & 0.49\\\hline
$P_{1}$ & 96.19 & \textit{0.021 } & 93.71 & 0.56\\\hline
$P_{2}$ & 93.26 & \textit{0.055 } & 92.98 & 0.77\\\hline
$P_{3}$ & 97.64 & \textit{0.0010 } & 97.27 & 0.78\\\hline
$P_{4}$ & 96.59 & \textit{0.0032 } & 95.68 & 0.72\\\hline
\end{tabular}

$^{a}\ $The operator observed at time $T_{m}=T_{f}/2$. Here $\mu$ is the
dipole; $H_{0}$ is the field-free

\ \ Hamiltonian; $P_{k}$ is a population projection operator for state
$\left\vert k\right\rangle $, $k=0,\cdots,4$.

$^{b}\ $Yield from the optimal control field and an instantaneous observation
at time $T_{m}=T_{f}/2$.

$^{c}\ $Observed value of operator $A$.

$^{d}\ $Yield arising from the control field without actually performing the
observation, but

\ \ the control field is determined in the presence of the observation of
operator $A$.

$^{e}\ $Fluence of the control field.\pagebreak

TABLE II. Optimal control fields interacting with an observation of the dipole
$\mu$ for model 1 with different objective yields.%

\begin{tabular}
[c]{|c|c|c|c|c|}\hline
$O_{T}$ $^{a}$(\%) & $O\left[  E\left(  t\right)  ,\mu\right]  ^{b}$(\%) &
$O\left[  E\left(  t\right)  ,0\right]  ^{c}$(\%) & $F^{d}$ & $F_{0}$ $^{e}%
$\\\hline
10 & 10.00 & 2.03$\times$10$^{-7}$ & 0.0020 & 0.017\\\hline
20 & 20.03 & 2.74$\times$10$^{-9}$ & 0.00026 & 0.023\\\hline
30 & 29.86 & 0.0052 & 0.0034 & 0.027\\\hline
40 & 39.82 & 2.69 & 0.017 & 0.031\\\hline
50 & 49.73 & 13.87 & 0.027 & 0.034\\\hline
60 & 59.73 & 40.20 & 0.036 & 0.038\\\hline
70 & 69.74 & 48.12 & 0.041 & 0.042\\\hline
80 & 79.18 & 81.80 & 0.31 & 0.046\\\hline
90 & 88.86 & 89.36 & 0.34 & 0.052\\\hline
100 & 92.42 & 94.03 & 0.37 & 0.063\\\hline
\end{tabular}

$^{a}$ Objective yield in Eq. (\ref{J0}).

$^{b}$ Yield from an optimal control field and an observation of the dipole
$\mu$ at time $T_{m}=T_{f}/2$.

$^{c}$ Yield arising from the control field without an observation of the
dipole, but with the

\ \ control field determined in the presence of an observation of the dipole.

$^{d}$ Fluence of the control field optimized with the observation present.

$^{e}$ Fluence of the control field optimized without the observation present.

\pagebreak

\bigskip TABLE III. Optimal control of model 2 with a sequence of
instantaneous observations%

\begin{tabular}
[c]{|c|c|c|c|}\hline
$N$ $^{a}$ & $O\left[  \mathbf{P}\right]  ^{b}$(\%) & $O\left[  E\left(
t\right)  ,\mathbf{P}\right]  ^{c}$(\%) & $O\left[  0,\mathbf{P}\right]  ^{d}%
$(\%)\\\hline
0 & 0 & 12.93$^{e}$ & 0.00\\\hline
1 & 50.00 & 56.46 & 11.82\\\hline
3 & 62.50 & 72.60 & 16.90\\\hline
5 & 71.04 & 79.22 & 20.46\\\hline
7 & 73.72 & 80.61 & 18.22\\\hline
9 & 80.11 & 80.45 & 19.65\\\hline
\end{tabular}

$^{a}$ Number of observations $N$ performed at times $T_{k}=\frac{k}{N+1}%
T_{f}$, $k=1,\cdots N$.

$^{b}$ Yield from the optimal observations without a control field.

$^{c}$ Yield from the optimal observations in the presence of a non-optimal
control field.

$^{d}$ Yield from the optimal observations without a control field, but with
the optimal

\ \ observations determined in the presence of non-optimal control field.

$^{e}$ The fluence of the non-optimal control field is $F=0.0196$.

\bigskip\pagebreak

\bigskip\bigskip TABLE IV. Optimal control of model 3 with various
instantaneous observations at time $T_{m}=T_{f}/2$.%

\begin{tabular}
[c]{|c|c|c|c|c|}\hline
$P\ ^{a}$ & $O\left[  E\left(  t\right)  ,P\right]  \ ^{b}$(\%) & Tr$\left[
\rho\left(  T_{m}\right)  P\right]  \ ^{c}$ & $O\left[  E\left(  t\right)
,0\right]  \ ^{d}$(\%) & $F\ ^{e}$\\\hline
-- & 49.99 & -- & 49.99 & 0.0031\\\hline
$P_{0}$ & 66.90 & 0.068 & 46.04 & 0.76\\\hline
$P_{1}$ & 49.99 & 0.50 & 50.00 & 0.96\\\hline
$P_{2}$ & 66.66 & 0.066 & 46.37 & 0.49\\\hline
\end{tabular}

$^{a}$\ The operator observed at time $T_{m}=T_{f}/2$. Here $P_{k}$ is a
population projection operator

\ \ for state $\left\vert k\right\rangle $, $k=0$, $1$, $2$.

$^{b,c,d,e}$ Refer to Table I.

\pagebreak

TABLE V. Control of model 3 with an optimized continuous observation.%

\begin{tabular}
[c]{|c|c|c|c|c|c|}\hline
$P\ ^{a}$ & $O\left[  A,\gamma,T_{1},T_{2}\right]  \ ^{b}$(\%) & $F\ ^{c}$ &
$\gamma$ & $T_{1}$ & $T_{2}$\\\hline
-- & 49.99 & 0.0031 & -- & -- & --\\\hline
$P_{0}$ & 98.92 & 0.021 & 5.00 & 119.13 & 199.96\\\hline
$P_{1}$ & 49.99 & 0.0086 & 0.0025 & 6.24 & 6.33\\\hline
$P_{2}$ & 99.55 & 0.0037 & 5.00 & 2.55 & 199.24\\\hline
\end{tabular}

$^{a}$ The population operator $P_{k}$ is observed between time $T_{1}$ and
$T_{2}$ with the strength $\gamma$.

\ \ Here $P_{k}$ indicates observation of the population in state $\left\vert
k\right\rangle $, $k=0$, $1$, $2$.

$^{b}$ Yield from the optimal control field and a continuous observation
between times $T_{1}$ and $T_{2}$ with the strength $\gamma$.

$^{c}$ Fluence of the control field.

\bigskip\bigskip\pagebreak

\bigskip

TABLE VI. Optimal control of Model 4 with different continuous quantum observations

\bigskip%
\begin{tabular}
[c]{|c|c|c|}\hline
$\kappa^{\ a}$ & $O\left[  E(t),P_{1^{\prime}}\right]  $(\%)$^{b}$ & $\left.
P_{1^{\prime}}\right.  ^{c}$\\\hline
0.00 & 71.96 & 14.22\\\hline
0.01 & 75.53 & 13.52\\\hline
0.03 & 80.77 & 12.03\\\hline
0.05 & 84.32 & 10.50\\\hline
0.09 & 88.61 & 8.27\\\hline
0.15 & 91.81 & 6.43\\\hline
0.20 & 93.28 & 5.33\\\hline
0.30 & 94.78 & 4.25\\\hline
\end{tabular}

$^{a}$ Observation strength of state $\left\vert 1^{\prime}\right\rangle $;
refer to Eq. (\ref{M4Dyna}).

$^{b}$ Population yield in the target state $\left\vert 3\right\rangle $ from
the optimal control field and continuous

\ \ observations of the population in state $\left\vert 1^{\prime
}\right\rangle $.

$^{c}$ Population in the undesired state $\left\vert 1^{\prime}\right\rangle $.

\bigskip\pagebreak

Fig\ref{Fig_Model}. \bigskip Three multilevel systems used to investigate the
impact of observations in the optimally controlled quantum dynamics
simulations in Sec. IV. (a) The five-level ladder configuration used for
models 1 and 2. (b) Model 3 with degenerate transition frequencies
$\omega_{01}=\omega_{02}$. (c) Model 4, where the two transition frequencies
$\omega_{11^{\prime}}=\omega_{23}$ are degenerate.

\begin{figure}
[h]
\begin{center}
\includegraphics[
trim=0.000000in 0.000000in 0.000000in -0.098417in,
height=3.0115in,
width=6.0231in
]%
{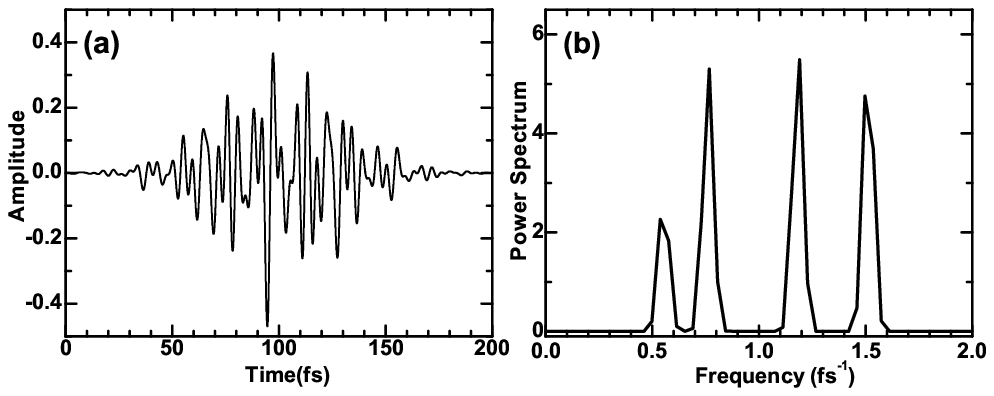}%
\caption{F Shuang et al}%
\label{Fig_Field}%
\end{center}
\end{figure}

Fig\ref{Fig_Field}. The optimal control field and its power spectrum for model
1 without an observation being present. The field is found using the cost
function in Eq. (\ref{J0}) with a high expected yield of $O_{T}=100\%$. The
spectral features are at the system transition frequencies.

\begin{figure}
[h]
\begin{center}
\includegraphics[
trim=0.000000in 0.000000in 0.000000in -0.866850in,
height=4.5164in,
width=6.0231in
]%
{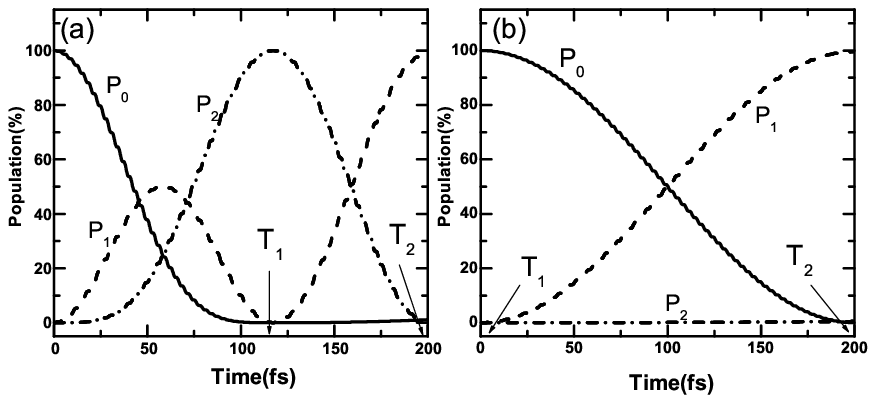}%
\caption{F Shuang et al}%
\label{Fig_Symm}%
\end{center}
\end{figure}

Fig\ref{Fig_Symm}. The population evolution of model 3 driven by an optimal
control field with the help of optimized continuous observations performed
between time $T_{1}$ and $T_{2}$. $P_{k}$ denotes the population in level $k$,
$k=1,2,3$. The observation is on state $\left\vert 0\right\rangle $ in plot
(a) and on state $\left\vert 2\right\rangle $ in plot (b).

\bigskip\bigskip\pagebreak

\bigskip%
\pagebreak%
\pagebreak\pagebreak%

\end{document}